\documentstyle[psfig]{aa}

\begin{document}

   \thesaurus{06         
              (03.11.1;  
               16.06.1;  
               19.06.1;  
               19.37.1;  
               19.53.1;  
               19.63.1)} 
   \title{A photometric study of the intermediate age open cluster 
King~5\thanks{Based on observations taken at TIRGO.}}

   \author{Giovanni Carraro 
          \inst{1} and  Antonella Vallenari\inst{2}}

   \offprints{Giovanni Carraro ({\tt carraro@pd.astro.it})}

   \institute{Dipartimento di Astronomia, Universit\'a di Padova,
	vicolo dell'Osservatorio 5, I-35122, Padova, Italy
        \and
        Osservatorio Astronomico di Padova, vicolo Osservatorio 
          5, I-35122, Padova,
	Italy\\
        e-mail: {\tt 
carraro,vallenari\char64pd.astro.it}
             }

   \date{Received ; accepted}

   \maketitle

   \markboth{Carraro et al}{King~5}

   \begin{abstract}

We report on near IR ($J$ and $K$ bands) observations of an $8^{\prime} 
\times 8^{\prime}
$ region centered on the poorly studied open cluster
King~5, for which only optical photometry 
existed. Photometry of two nearby fields is also reported.
We found that the cluster is of moderate age ($1.0~Gyr$ old), intermediate
in age between the Hyades and NGC~752. Combining optical and IR photometry
we obtain estimates for the cluster parameters. 
The color excesses E(J-K), E(V-I) and E(V-K) 
are 0.50, 1.10 and 2.45, respectively. The true distance modulus
turns out to be $(m-M)_o~=~11.40\pm 0.15$. As a consequence, King~5 is 1.9~kpc
far from the Sun.

      \keywords{Photometry~:~Infrared--Open clusters and associations
	  ~:~King~5~:~individual. 
               }
   \end{abstract}

%

\section{Introduction}
This paper continues a series dedicated to the presentation of near-infrared 
photometry (in $J$ and $K$ pass-bands) for northern galactic open clusters.
We already reported on the very young open clusters NGC~1893 and Berkeley~86 
(Vallenari et al 1999a), the old clusters Berkeley~17 and Berkeley~18 
(Carraro et al 1999a), and the intermediate age clusters
IC~166 and NGC~7789 (Vallenari et al 1999b). 
Here we present $J$ and $K$ photometry for King~5,
a poorly studied open cluster for which no detailed study has been reported
so far.\\
King~5 is located near the galactic plane 
at $l~=~143^{o}.75$ and $b~=-4^{o}.27~$, and it is designated also as OCL~384
and C0311+525 by IAU. Its diameter is estimated to be 
about $5^{\prime}-8^{\prime}$.\\
A preliminary investigation was conducted by Phelps et al (1994) who obtained
VI photometry for 1343 stars in a region of $11^{\prime} \times 11^{\prime}$.
This study revealed that King~5 is a cluster of the Hyades generation, with 
an age around one billion years, derived from a morphological age estimator.\\
These authors  do not report estimates for cluster distance and reddening.\\
Moderate resolution spectra for 24 stars in the field of King~5 have been 
obtained by Scott et al (1995). From this analysis it emerges that King~5
belongs to the old thin disk population, and its rotational velocity around 
the galactic center ($-52~km/sec$) is consistent with Hron (1987) rotation curve
at the cluster location.\\
On the base of the radial velocities 19 members have been singled out.\\
More recently Durgapal et al (1998) published UBVRI photometry of King~5,
inferring by comparison with solar metallicity
isochrones an age slightly lower than 1~Gyr, a distance 
of 2~kpc and a reddening E(V-I)=0.94.\\
Finally it is worth recalling that the metal abundance of King~5 has been
determined by Friel et al (1995) using medium resolution spectroscopy of 
twelve giant stars. From this study $[Fe/H]$ turns out to be ($-0.38\pm0.20$),
half than the solar value.\\
In this paper we combine together IR and optical photometry to obtain
an estimate of the cluster fundamental parameters.
We find that the cluster is $1.0~Gyr$ old.\\

The plan of the paper is as follows: Section~2 is devoted to the presentation
of data acquisition and reduction; in Section~3 we discuss the morphology
of the Color Magnitude Diagrams (CMDs) for different pass-bands;
Section~4 is dedicated to the derivation of the color excess, while
age and distance are inferred in Section~5. Finally our conclusions
are summarized in Section~6.

\begin{table*}
\caption[ ]{ Observation Log Book }
\tabcolsep 0.8cm
\begin{tabular}{c|c|c|c|c|c}
\hline
\hline
Cluster               &$\alpha$    &$\delta$  & Date& \multicolumn{2}{c}
{Exposure Times (sec)} \\ 
                           &(2000)      &(2000) &    &J&K\\
\hline
King~5   & 3 14 47& 52 42 39& Oct, 25, 1997& 540 & 880\\
King~5 F1& 3 14 23& 52 48 39& Oct, 27, 1997& 480 & 620\\
King~5 F2& 3 15 42& 52 33 17& Oct, 27, 1997& 480 & 620\\
K5A      & 3 14 40& 52 42 51& Oct, 25, 1997&     &    \\
K5B      & 3 14 27& 52 42 02& Oct, 25, 1997&     &    \\
\hline
\hline
\end{tabular}
\end{table*}

\begin{figure*}
\centerline{\psfig{file=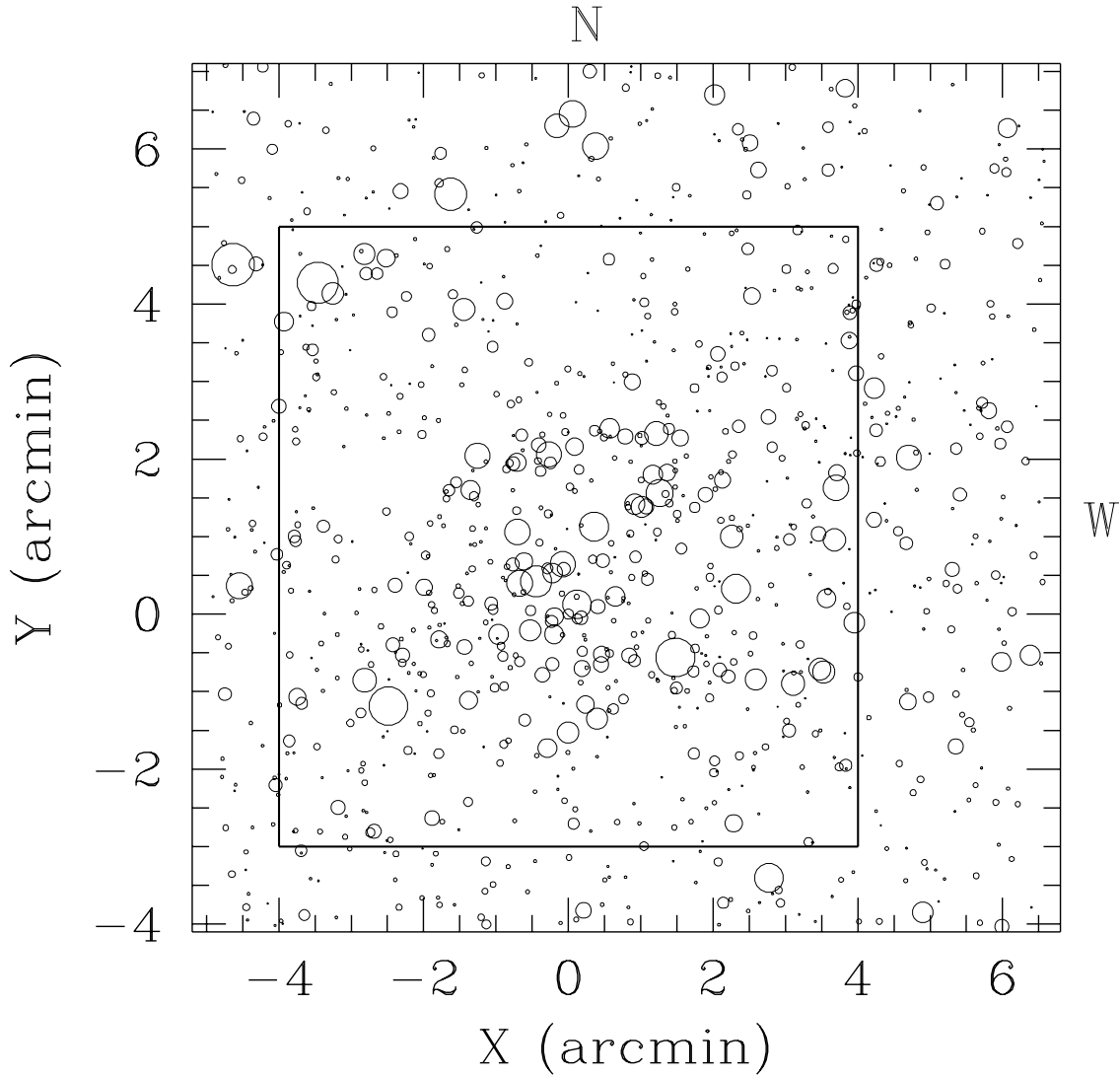,width=16cm,height=16cm}}
\caption{The field covered in the region of King~5 by the optical
photometry of Phelps et al 1994. The inner region shows the mosaic
of the four fields observed by us in the IR. North is up, East on the left.}
\end{figure*}

\begin{figure*}
\centerline{\psfig{file=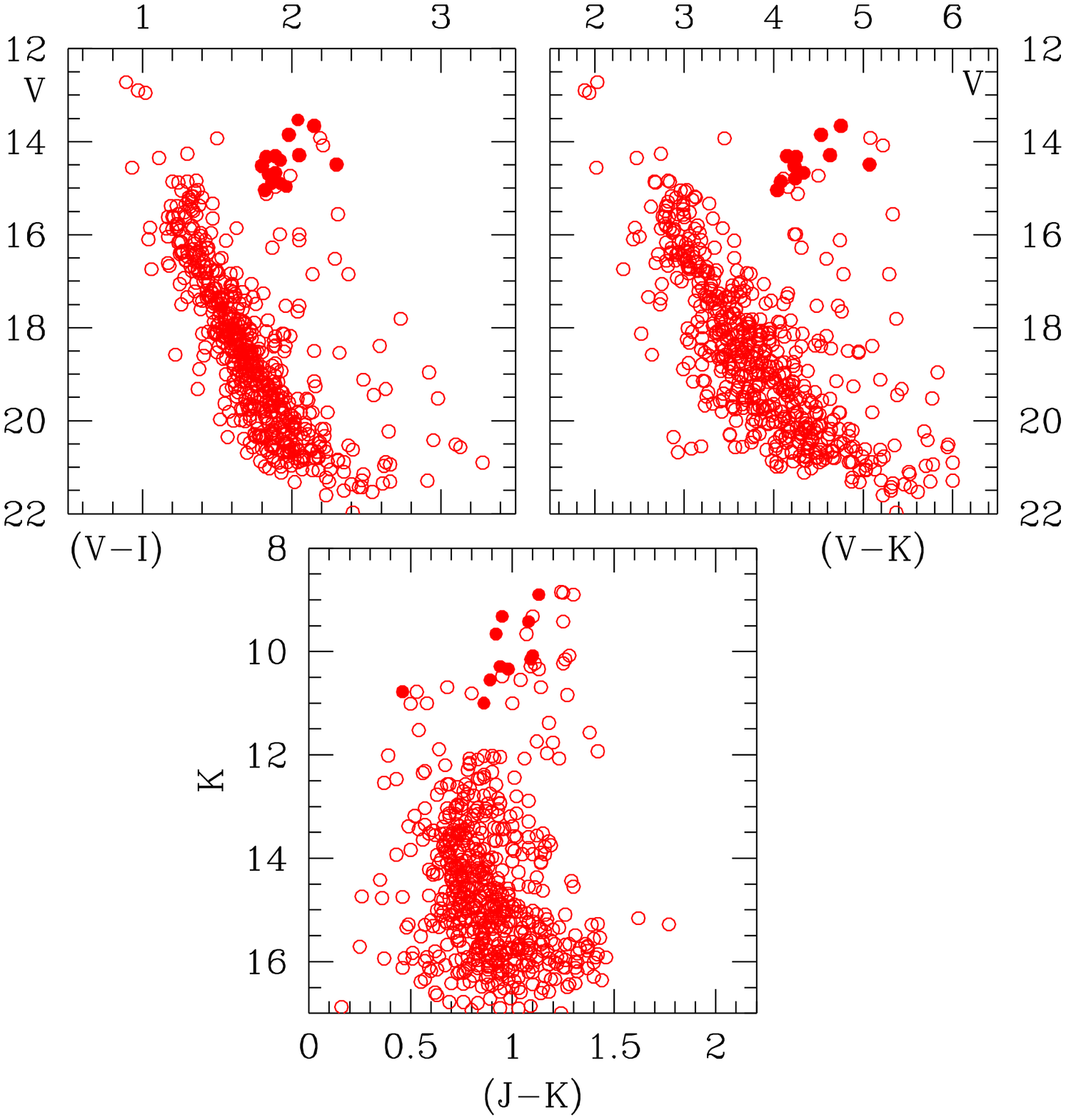,width=16cm,height=16cm}}
\caption{The CMD of King~5. The optical
photometry (V-(V-I), upper left panel) is taken from Phelps et al (1994).
The IR photometry (lower panel) is that presented in this paper,
whereas the upper right panel present the derived CMD in V-(V-K).
Filled circles represent cluster
members on the base of the radial velocity.}
\end{figure*}

\section {Observations and Data reduction}

J (1.2 $\mu$m) and K (2.2 $\mu$m) photometry of King~5
was obtained 
at the  1.5m Gornergrat Infrared Telescope (TIRGO) 
equipped with Arcetri Near Infrared Camera (ARNICA)
in October 1997. 
ARNICA is based  on a NICMOS3 256$\times$ 256
pixels array (gain=20 e$^-/$ADU, read-out noise=50 e$^-$,
 angular scale =1$\arcsec/$pixel, and $4 \times 4 \arcmin^{2}$ field of
view). 
More details about the observational equipment and infrared camera,
and the reduction procedure can be found in Carraro et al (1999a).
Through each filter 4 partially overlapping images of
each field were obtained,
 covering a total field of view of about 8 $\times 8 \arcmin^{2}$,
in short exposures to avoid sky saturation.
One field is located close to the center of King~5.
In addition, a sample of the background population of King~5 was 
obtained observing a field of $2^{\prime} \times 2^{\prime}$ 
at about $15^{\prime}$ South-East from the cluster center
and a second one of the same dimensions located at about $5^{\prime}$ 
Northward.

 The log-book of the observations is presented
 in Table~1 where the centers of the observed fields and the total
exposure times are given.
 The nights were photometric
with a seeing of 1$\arcsec$-1.5$\arcsec$. 
Fig.1 presents the final mosaic of the 4 frames
for King~5 in K passband.

\begin{figure*}
\centerline{\psfig{file=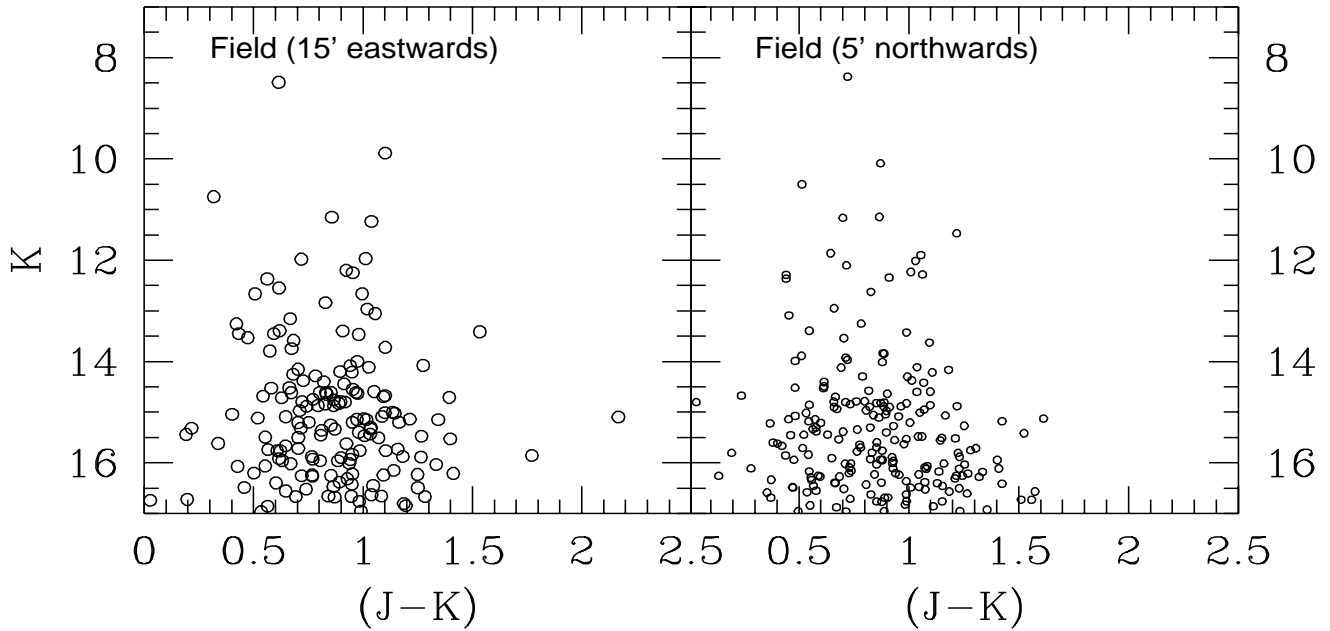,width=18cm,height=17cm}}
\caption{The CMD of two comparison fields taken $15^{\prime}$ eastwards 
(left panel) and $5^{\prime}$ northwards (right panel) of King~5.}
\end{figure*}

The conversion of the
instrumental magnitude j and k to the standard
J, K was made using stellar fields of standard stars taken
from   Hunt et al (1998) list.
About 10 standard stars per night have been used. In addition 2 stars 
(thereafter K5A and K5B) inside the field have been measured with
a photometer and used for the calibration. Results are summarized in Table~2.

\begin{table}
\caption{Photometry of K5A and K5B}.
\tabcolsep 0.6cm
\begin{tabular}{ccc}
\hline
\hline
\multicolumn{1}{c}{Star} &
\multicolumn{1}{c}{$K$}  &
\multicolumn{1}{c}{$J$}  \\
\hline
K5A    & 9.20$\pm$0.04 & 10.41$\pm$0.05 \\
K5B    & 8.74$\pm$0.04 & 10.10$\pm$0.05 \\
\hline
\hline
\end{tabular}
\end{table}

The  relations in usage per 1 sec exposure time are: 
\begin{equation}
J  = j+19.51
\end{equation}
\begin{equation}
K  = k+18.94
\end{equation}
with standard deviation of the zero points of 0.03  mag for the $J$ 
and 0.04 for the $K$ magnitude. This error is only due to the linear
interpolation of the standard stars. The calibration uncertainty
is dominated by the error due to the correction from aperture photometry
to PSF fitting magnitude. 
The standard stars used for the calibration do  not cover the entire
colour range of the data, because of the lack of stars redder than
$(J-K) \sim 0.8$. From our data, no colour term is found for $K$ mag,
whereas we cannot exclude 
it  for the $J$ magnitude.
Taking all into account,
we estimate that the total error
on the calibration is about 0.1 mag both in $J$ and $K$ pass-bands.

\section{The Color Magnitude Diagrams}
The CMD derived from optical photometry is shown in the upper left panel
of Fig.~2.
The global morphology resembles the CMD of an intermediate age open cluster
like NGC~2477 (Hartwick et al 1972), Tombaugh~1 (Carraro \& Patat 1995)
and NGC~6603 (Bica et al 1993). The MS extends vertically down to $V \approx
22$, and the Turn Off point (TO) is situated at $V~=~15.5$, $(V-I)~=~1.3$.
The brightest stars in the red region of the diagram 
can represent the Red Giant (RG) clump
of core He-burning stars. Most of these are cluster members, on the base
of their radial velocities (Scott et al 1995). The Herzsprung gap is clearly
defined.
The population of stars on the right side of the MS are probably interlopers,
showing that the CMD is contaminated by stars
from the field of the Galactic Disk. Also the stars above the TO 
probably belong to the field.\\

The CMD from IR photometry is shown in the lower panel of Fig.~2, and appears
significantly different. In fact it is quite similar to the CMD of old
open cluster like M~67 (Montgomery et al 1993) and  Berkeley~39
(Kassis et al 1997). The MS extends down to $K~=~16.5$, and the TO is
located at $K~=~12.5$, $(J-K)~=~0.70$. The region above the TO 
might contain some interlopers (see Fig.~3).\\
Apparently there is no Herzsprung gap,
and the RGB is sparsely populated, like in all clusters with age
greater that $4~Gyr$. The stars at $K \approx 10.3$, $(J-K) \approx 1.1$
might represent the clump of core He burners, quite small in this
age range. We recall that IR photometry covers all the cluster,
and the foreground contamination is negligible, as it can be seen in Fig.~3,
where two comparison fields $15^{\prime}$ eastwards 
and $5^{\prime}$ northwards, respectively, are shown.\\

\begin{figure}
\centerline{\psfig{file=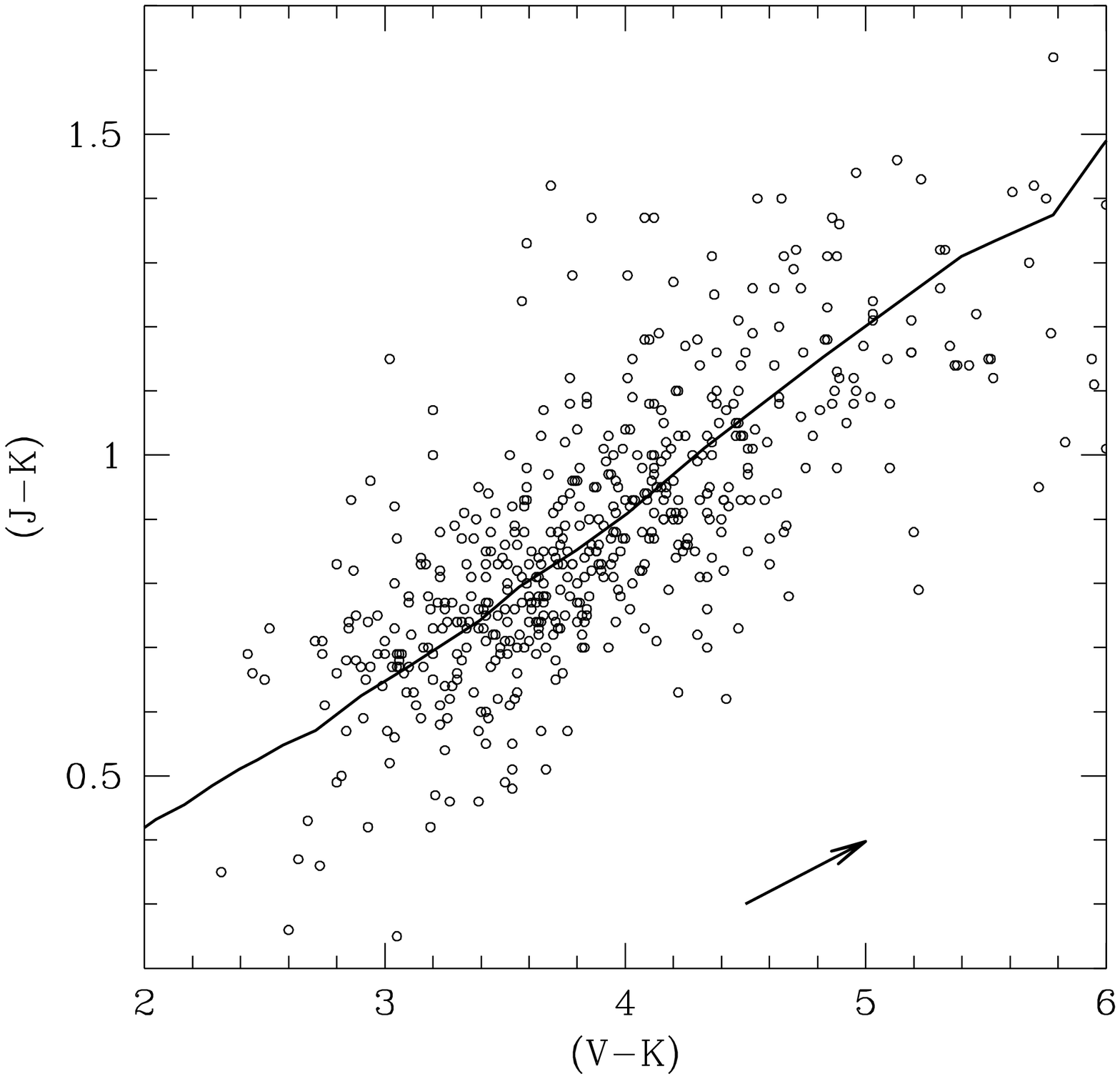,width=9cm,height=9cm}}
\caption{Two colors diagram for MS stars in King~5. The solid line is a Zero
Age MS for $[Fe/H]~=~-0.38$. In the lower left corner the reddening vector is shown.
See text for any detail.}
\end{figure}

\begin{figure}
\centerline{\psfig{file=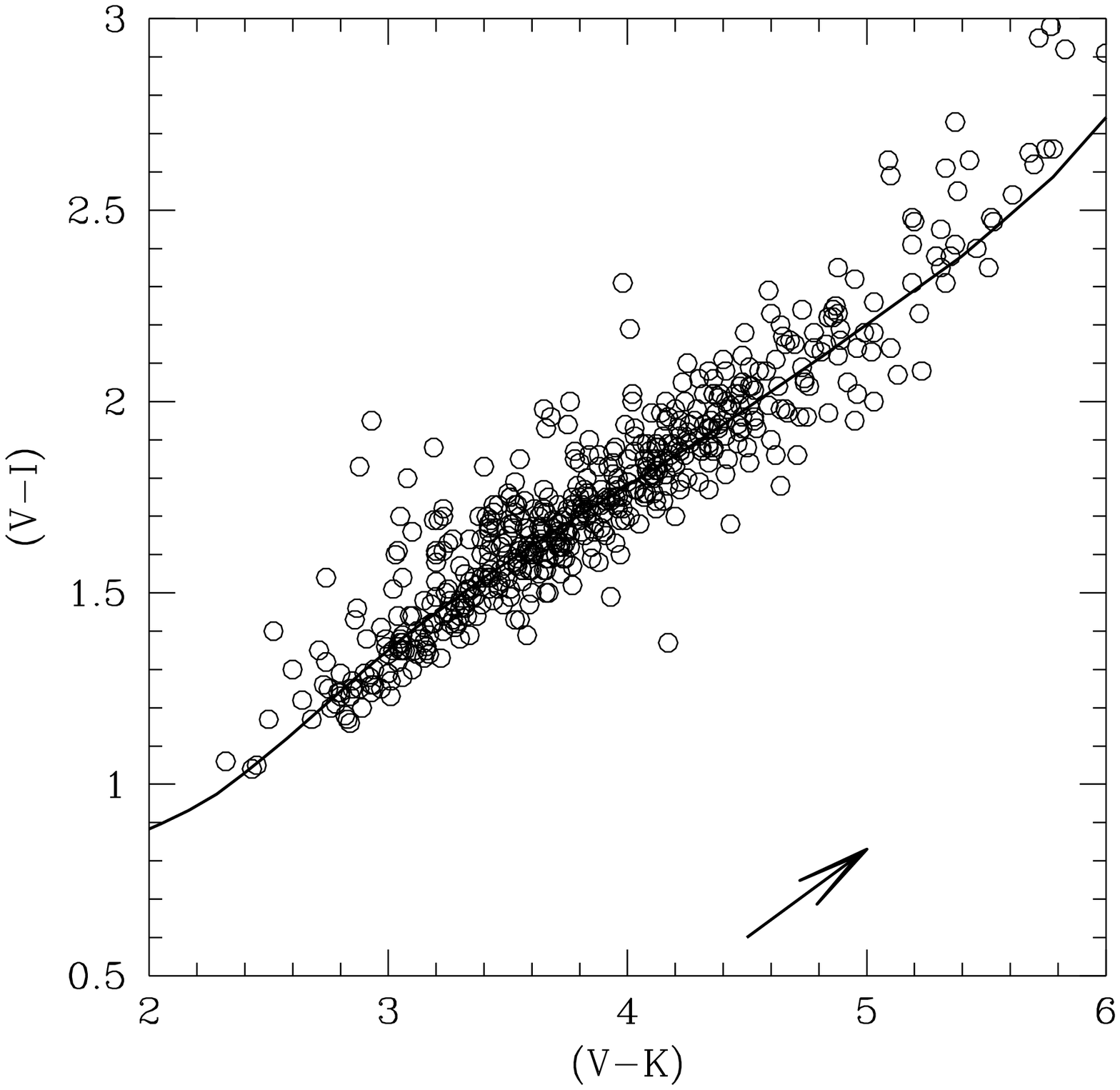,width=9cm,height=9cm}}
\caption{Two colors diagram for MS stars in King~5. The solid line is a Zero
Age MS for $[Fe/H]~=~-0.38$. In the lower left corner the reddening vector is shown.
See text for any detail.}
\end{figure}

In order to clarify the nature and age range of this cluster,
and whether the differences are simply due to the use of different
colors, we have combined
optical and IR data to construct the CMD in $V - (V-K)$, which is shown 
in the lower right panel Fig.~2. This CMD looks like quite similar to the optical
CMD shown in the same figure,
demonstrating that King~5 is an intermediate age open cluster.
To better assess this point, in the next sections we are going to determine
in details the cluster basic parameters.

\begin{figure*}
\centerline{\psfig{file=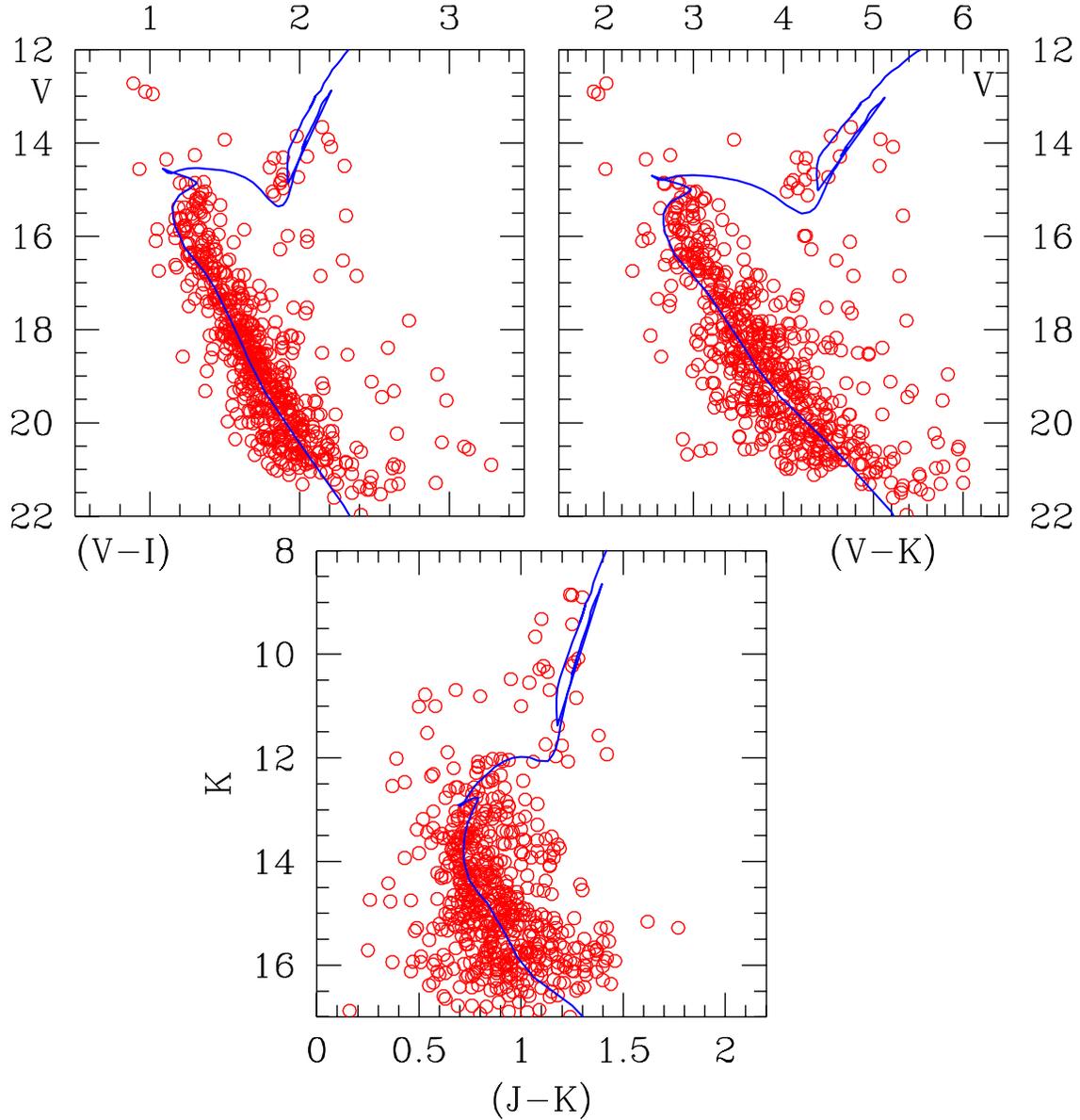,width=16cm,height=16cm}}
\caption{Age determination for King~5. Overimposed on the CMDs
is a $Z~=~0.008$ isochrone for an age of $1~Gyr$. Lower panel shows
the fit in the plane $K -(J-K)$, upper left panel in the plane $V -(V-I)$,
and upper right panel in the plane $V -(V-K)$.
See the text for any detail.}
\end{figure*}

\section{Reddening}
In order to derive the interstellar extinction for King~5, we have combined
optical and IR photometry ($600$ stars in total), and construct two color 
diagrams, namely $(J-K)~vs~(V-K)$ and $(V-I)~vs~(V-K)$, which are shown in Fig.~4 
and Fig.~5, respectively. Only MS stars are considered.
We superimposed a Zero Age MS (ZAMS) for the theoretical
metal content $Z~=~0.008$, obtained translating the observed $[Fe/H]$ by means
of the relation:

\[
[Fe/H] = \log \frac{Z}{0.019}
\]

\noindent
taken from Bertelli et al (1994).
Since the cluster does not exhibit a RGB, it is not possible to derive
an independent photometric estimate for the metal content.\\
In Fig.~4 the fit has been obtained shifting the ZAMS with
E(J-K)~=~0.50 and E(V-K)~=~2.45, which corresponds to a ratio 
$\frac{E(V-K)}{E(J-K)}~=~4.9$, relatively close to the value $5.3$
suggested by Cardelli et al (1989).\\
The fit in Fig.~5, on the other hand, has been achieved shifting the ZAMS by
E(V-I)~=~1.10 and E(V-K)~=~2.45, whose ratio turns out to
be $\frac{E_{(V-I)}}{E_{(V-K)}}~=~0.44$, in agreement with the value $0.42$
from Cardelli et al (1989).
Although reasonable, these estimates are affected by the limitation that
the reddening vector is almost parallel to the ZAMS.  However, at least
for E(V-I), our results do not differ to much from Durgapal et al (1998),
although these author did not use the spectroscopic metallicity from
Friel et al (1995). Indeed, by using a lower metallicity, one should expect a
larger reddening.

\section{Age and distance}
The knowledge of metallicity and reddening, allows us to infer
the distance and age of King~5 by means of fitting with isochrones
(Bertelli et al 1994).
The fit is shown in the three panels of 
Fig.~6, for the planes $V$ vs $(V-I)$, $V$ vs $(V-K)$, and $K$ vs $(J-K)$, 
respectively.\\
We have adopted the theoretical metallicity $Z~=~0.008$ above derived,
and performed the fitting with a $1.0~Gyr$ isochrone, which better
matches the observational data.
The criterion that guided us in performing this comparison was the simultaneous
fit of the TO and the clump magnitudes. Since no membership exists for this
cluster, it is not possible to define exactly the MS TO, which is populated also
by unresolved binary stars, and interlopers (see the field population in Fig.~3
for an interpretation of the TO region in the IR CMD).
However the clump luminosity can be reliably determined, since many stars
are also spectroscopic members.\\
By using the reddening estimated derived in the previous Section,
the apparent distance moduli $(m-M)_{K,(J-K)}$ and $(m-M)_{V,(V-I)}$,  
in the plots turn out to be
$11.90$ and $14.00$, respectively. The latter values can also be obtained
assuming that the mean clump magnitude is $M_{V}~=~0.80$ (Girardi et al 1998).
Once corrected, these
values converge to the absolute distance modulus $(m-M)_o~=~11.40 \pm 0.15$.\\
King~5 turns out to be 1.9~kpc distant from the Sun, and about 9.6~kpc
far from the Galactic Center.

\section{Conclusions}
In this paper we have presented a detailed study of the poorly
known intermediate age open cluster King~5.
By combining optical and IR photometry we have proved that King~5 is a
moderate age open cluster about $1~Gyr$ old, intermediate in age between the 
Hyades and NGC~752.\\
BY studing the CMD and two color digrams,
we have obtained estimates for the cluster reddening
and distance. In detail, we found that the color excesses 
$E_{(J-K)}$, $E_{(V-I)}$ and $E_{(V-K)}$ 
are 0.50, 1.10 and 2.45, respectively, and that their ratios are in agreement
with the standard values (Cardelli et al 1989).
The derived corrected distance modulus $(m-M)_o~=~11.40$ implies
a distance of 1.9~kpc from the Sun.

\begin{acknowledgements}
G. C. acknowledges useful e-discussions with Prof. Ken Janes,
and the anonymous referee who helped to improve the presentation of the paper.
This study has been financed by the Italian Ministry of
University, Scientific Research and Technology (MURST) and the Italian
Space Agency (ASI).
\end{acknowledgements}

{}

\end{document}